\documentclass[twocolumn,showpacs,preprintnumbers,amsmath,amssymb]{revtex4}

\usepackage{graphicx}
\usepackage{dcolumn}
\usepackage{bm}

\begin{document}


\title{Approximate action-angle variables for the figure-eight
and other periodic three-body orbits}

\author{Milovan \v Suvakov}
\email{suki@ipb.ac.rs}

\affiliation{Institute of Physics, Belgrade
University, Pregrevica 118, Zemun, \\
P.O.Box 57, 11080 Beograd, Serbia}%

\author{V. Dmitra\v sinovi\' c}
\email{dmitrasin@ipb.ac.rs}

\affiliation{Yukawa Institute for Theoretical Physics, Kyoto University, \\
Kyoto 606-8502, Japan, \\
Permanent address: Institute of Physics, Belgrade
University, Pregrevica 118, Zemun, \\
P.O.Box 57, 11080 Beograd, Serbia }

\date{\today}

\begin{abstract}
We use the maximally permutation symmetric set of three-body
coordinates, that consist of the ``hyper-radius" $R =
\sqrt{\rho^{2} + \lambda^{2}}$, the ``rescaled area of the
triangle" $\frac{\sqrt 3}{2 R^2} |{\bm \rho} \times {\bm
\lambda}|$) and the (braiding) hyper-angle $\phi =
\arctan\left(\frac{2 {\bm \rho} \cdot {\bm \lambda}}{\lambda^2 -
\rho^2} \right)$, to analyze the ``figure-eight" choreographic
three-body motion discovered by Moore \cite{Moore1993} in the
Newtonian three-body problem. Here ${\bm \rho}, {\bm \lambda}$ are
the two Jacobi relative coordinate vectors. We show that the
periodicity of this motion is closely related to the braiding
hyper-angle $\phi$. We construct an approximate integral of motion
${\overline{G}}$ that together with the hyper-angle $\phi$ forms
the action-angle pair of variables for this problem and show that
it is the underlying cause of figure-eight motion's stability. We
construct figure-eight orbits in two other attractive
permutation-symmetric three-body potentials. We compare the
figure-eight orbits in these three potentials and discuss their
generic features, as well as their differences. We apply these
variables to two new periodic, but non-choreographic orbits: One
has a continuously rising $\phi$ in time $t$, just like the
figure-eight motion, but with a different, more complex
periodicity, whereas the other one has an oscillating $\phi(t)$
temporal behavior.
\end{abstract}


\maketitle

\section{Introduction}
\label{intro}

The three-body problem is one of the oldest and most challenging
in classical mechanics \cite{Valtonen:2005}. Until recently only a
few periodic three-body solutions were known \cite{Valtonen:2005}
in Newton's gravitational interaction potential. A new periodic,
``figure eight", trajectory was found in 1993 by Moore
\cite{Moore1993} in the case of three equal masses and
gravitational $-1/r$ potential, using numerical methods. Its
existence and stability were later proven formally by way of
variational arguments \cite{Chenciner2000}, but no closed
(analytic) form of this solution has been shown as yet. Moreover,
the figure-eight solution has also been found in the
general-relativistic three-body dynamics \cite{Imai2007}, and its
bifurcations have been studied as a function of the mass asymmetry
\cite{Galán2002}. Proofs of existence, as well as some properties
of figure-eight orbits in pairwise sums of $-1/r^{\alpha}$
two-body potentials with $\alpha \neq 1$, have been studied in
Refs. \cite{Fujiwara2003a}, \cite{Fujiwara2003b},
\cite{Fujiwara2004}. Any new solution, and/or insight into the
existing ones should be of intrinsic interest.

Of course, the figure-eight orbit is highly symmetric, but it is
not immediately clear what is the underlying dynamical reason for
this symmetry. It is an empirical fact, however, that all known
figure-eight orbits exist only in (three-body) permutation
symmetric potentials. Indeed, it is known that the figure-eight
orbit bifurcates into new, less symmetric orbits as one changes
the mass ratio(s) of the three particles, and thus breaks the
permutation symmetry, see Refs. \cite{Galán2002},
\cite{Broucke2006}. We explore this connection between the
permutation symmetry and the figure-eight orbit and make it more
explicit. In the process we have found new solutions with lesser
symmetry, much like those in Ref. \cite{Broucke2006}, and obtained
new insights into the role of permutation symmetry in the
classical three-body problem.

In this paper we report our studies of figure-eight orbits in
three kinds of three-body potentials: 1) the Newtonian gravity,
i.e., the pairwise sum of $-1/r$ two-body potentials; 2) the
pairwise sum of linearly rising $r$ two-body potentials (a.k.a.
the $\Delta$ string potential); 3) the Y-junction string potential
\cite{Dmitrasinovic:2009dy},\cite{taka01} that contains both a
genuine three-body part, as well as two-body contributions (this
is the first time that the figure-eight has been found in these
string potentials, to our knowledge). These three potentials share
two common features, {\it viz.} they are attractive and symmetric
under permutations of any two, or three particles \footnote{The
Coulomb interaction among three identical charged particles is
permutation symmetric, but repulsive.}.

A set of variables makes this permutation symmetry manifest and we
use them to plot the motion of a numerically calculated
figure-eight orbit. As there are three independent three-body
variables, and there can be at most two independent
permutation-symmetric three-body variables\footnote{there are two
irreducible one-dimensional representations of the permutation
group $s_3$.}, the third variable cannot be permutation-symmetric.
In other words the third variable must change under permutations.
Moreover, it must be a continuous variable and not be restricted
only to a discrete set of points, as is natural for permutations.
Thus it must provide a smooth interpolation between (discrete)
permutations. We identify here the third independent variable as
$\phi = \arctan\left(\frac{2 {\bm \rho} \cdot {\bm
\lambda}}{\lambda^2 - \rho^2} \right)$ and show that it
grows/descends (almost) linearly with the time $t$ spent on the
figure-eight trajectory and reaches $\pm 2\pi$ after one period
$T$. Thus, $\phi$ is, for most practical purposes, interchangeable
with the time variable $t$ on the figure-eight orbit. The
hyper-angle $\phi$ is the continuous braiding variable that
interpolates smoothly between permutations and thus plays a
fundamental role in the braiding symmetry of the figure-eight
orbits \cite{Moore1993,Rota1997}.

Then we construct the hyper-angular momentum $G_3 = \frac12
\left({\bf p_{\rho}} \cdot {\bm \lambda} - {\bf p_{\lambda}} \cdot
{\bm \rho} \right)$ conjugate to $\phi$, the two forming an
(approximate) pair of action-angle variables for this periodic
motion. Here we calculate numerically and plot the temporal
variation of $\phi$, as well as that of the hyper-angular momentum
$G_3(t)$, the hyper-radius $R(t)$ and $r(t)$. We show that the
hyper-radius $R(t)$ oscillates about its average value
$\overline{R}$ with the same angular frequency ($3 \phi$) and
phase, as the new (``reduced area") variable $r(t)$. Thus, we show
that $\phi(t)$ is, for most practical purposes, interchangeable
with the time variable $t$, in agreement with the tacit
assumption(s) made in Refs. \cite{Fujiwara2003a},
\cite{Chenciner2000}, though the degree of linearity of this
relationship depends on the precise functional form of the
three-body potential, see Sect. \ref{s:hyperangular dependence}.

As stated above, $\phi$ is not exactly proportional to time $t$,
but contains some non-linearities that depend on the specifics of
the three-body potential; consequently the hyper-angular momentum
$G_3$ is not an exact constant of this motion, but oscillates
about the average value $\overline{G}_3$, with the same basic
frequency $3 \phi$. Thus, the time-averaged hyper-angular momentum
$\overline{G}_3$ is the action variable conjugate to the
linearized hyper-angle $\phi^{'}$.

We use these insights to characterize two new planar periodic, but
not choreographic three-body motions with vanishing total angular
momentum. One of these orbits corresponds to a modification of the
figure-eight orbit with $\phi(t)$ that also grows more or less
linearly in time, but has a more complicated periodicity pattern
defined by the zeros of the area of the triangle formed by the
three particles (also known as ``eclipses", ``conjunctions" or
``syzygies"). Another new orbit has $\phi(t)$ that grows in time
up to a point, then stops and ``swings back". We show that this
motion, and the other two, can be understood in view of the
analogy between the three-body hyper-angular (``shape space")
Hamiltonian on one hand and a variable-length pendulum in an
azimuthally periodic in-homogeneous gravitational field, on the
other.

This paper is divided into five parts: after the Introduction in
Sect. \ref{s:perm coord} we introduce a complete (maximal) set of
permutation symmetric three-body variables and illustrate them
with two examples: 1) the curves in the ``shape space" of
triangles depicting those triangles with one of its three angles
equal to a particular value in the range $(\frac{\pi}{3}, \pi)$;
2) the contour plots of the Newtonian gravity, the Y-junction
string and the $\Delta$-string potentials. In Sect. \ref{s:figure
8} we show the time dependence of the figure-eight motion in
Newton's gravity and the Y-string potentials. In Sect. \ref{s:new
sol} we show and discuss the new solutions. Finally in Sect.
\ref{s:concl} we summarize and draw conclusions.

\section{Permutation symmetric three-body coordinates}
\label{s:perm coord}

As the static three-body potential depends on three independent
scalar variables, e.g. the pairwise relative
distances/separations, the choice of appropriate (relative)
variables is a crucial one. A number of three-body relative
variables have been devised, starting with those introduced by
C.G. Jacobi in the 19th century \cite{Whittaker:1993zz}, and
extending to the so-called hyper-spherical coordinates introduced
in the 1960's \cite{Smith:1959zz}, \cite{DELVES:1958zz},
\cite{Simonov:1965ei}. These variables were introduced in attempts
at solving certain quantum mechanical three-body problems, that
demand special attention to be paid to the permutation symmetry.
Nevertheless, only one, Ref. \cite{Smith:1959zz}, of these sets is
manifestly permutation symmetric and yet it has not been widely
used.

Here we use the manifestly permutation symmetric three-body
variables, apparently first introduced by Hopf: the hyper-radius
$R$, the ``scale-invariant area" of the triangle $\sqrt{1 - r^2} =
2 R^{-2}|{\bm \rho} \times {\bm \lambda}|$, where, and find as the
hyper-angle $\phi = \arctan\left(\frac{2 {\bm \rho} \cdot {\bm
\lambda}}{\lambda^2 - \rho^2} \right)$, that is conjugate to the
generalized hyper-angular momentum $G_3 = \frac12 \left({\bf
p_{\rho}} \cdot {\bm \lambda} - {\bf p_{\lambda}} \cdot {\bm \rho}
\right)$. One may relate these to the hyper-spherical variables
$x^{'} = \frac{2 {\bm \rho} \cdot {\bm \lambda}}{R^2}$ and $z^{'}
= \frac{\lambda^2 - \rho^2}{R^2}$ that have the circle with unit
radius as their natural domain. Then the area of the triangle
$\frac{\sqrt 3}{2} |{\bm \rho} \times {\bm \lambda}|$ and the
hyper-radius $R$ are related to the the new variables $r$, $\phi$
as follows
\begin{eqnarray}
r^2 &=& \, \left(x^{'2} + \, z^{'2} \right)
= \,1 - \,\left(\frac{2|{\bm \rho} \times {\bm \lambda}|}{R^{2}}\right)^{2} \\
\phi &=& \tan^{-1} \left(\frac{x^{'}}{z^{'}}\right).\
\end{eqnarray}
The hyper-angle $\phi$ is zero at the $(x=0,z=1)$ point (``12
o'clock") and increases as one moves clockwise.

\subsection{The shape space of triangles}
\label{s:shape space}

\begin{figure}[tbp]
\centerline{\includegraphics[width=2.5in,,keepaspectratio]{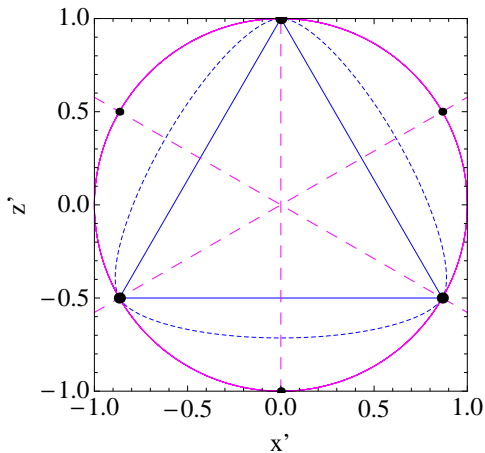}}
\caption{(Color online) The curves/lines in shape space of
triangles depicting triangles with one fixed angle: a) the outer
blue (dashed) line, for the fixed angle equal to $109.5^o$,
b) blue (full straight) lines for the fixed angle equal to
$\frac{\pi}{2}$, as functions of $z^{'}=z = \cos 2 \chi$ (ordinate
= vertical axis) and $x^{'} = x \sqrt{1-z^2} = \cos\theta
\sin2\chi$ (abscissa = horizontal axis). The domain of these
variables is a magenta (dark gray) circle of radius unity. The two 
straight red (dashed) lines at angles of $\pm \frac{2 \pi}{3}$, and the
vertical axis are the symmetry axes, i.e. $s_2$ subgroups of the
$s_3$ permutation group, and of the ``constant angle curves" in
shape space, as well. The three collinear configurations in which
one pair of particles has vanishing separation are denoted by big
solid circles, and the three collinear configurations in which one
particle has equal separations from the other two are denoted by
small solid circles.} \label{f:Combined_contour}
\end{figure}

The natural domain of the permutation symmetric variables is a
circle with unit radius, see Fig. \ref{f:Combined_contour}. The
points on the unit circle correspond to collinear configurations
(``triangles" with zero area).

The two straight lines at angles of $\pm \frac{2 \pi}{3}$,
together with the vertical axis are the three (reflection)
symmetry axes; these reflections correspond to the three
``two-body permutations"/transpositions in the $s_3$ permutation
group. The two cyclic permutations of the $s_3$ permutation group
correspond to the rotations through $\pm \frac{2 \pi}{3}$.

The six points where the symmetry axes cross the big circle in
Fig. \ref{f:Combined_contour} correspond to either a) three
collinear configurations (``shapes") in which one pair of
particles has vanishing separation (big solid circles), i.e.
``sits on top of each other", or b) three collinear configurations
(``shapes") in which one particle has equal separation from the
other two, i.e. ``sits in the middle between the other two" (small
solid circles). The center of the circle corresponds to the
equilateral triangle configuration (``shape"), which turns into a
point when the hyper-radius $R \to 0$.

\subsection{Newton's, $\Delta$ and Y-string potentials}
\label{s:Delta Y}

\begin{figure}[tbp]
\centerline{\includegraphics[width=2.5in,,keepaspectratio]{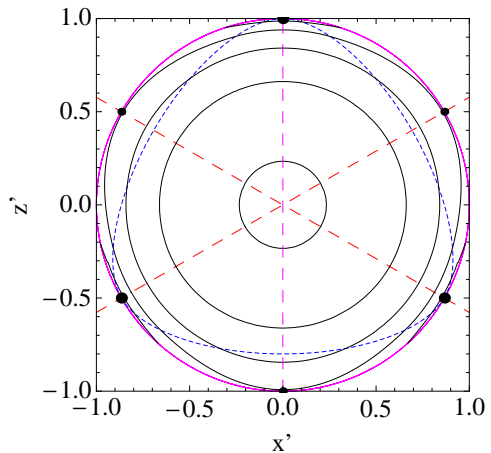}}
\caption{(Color online) The equipotential contours for the central
Y-string potential, and the boundary between the central Y-string
and two-string potentials as functions of $z^{'} = z = \cos 2
\chi$ (vertical axis), and $x^{'}=x \sqrt{1-z^2} = \cos
\theta \sin 2 \chi$ (horizontal axis). The blue
(dashed) curve denotes the boundary between the two-body and the
three-body components of this potential, see Ref.
\cite{Dmitrasinovic:2009dy}. The rotation symmetry about the axis
pointing out of the plane of the figure should be visible to the
naked eye.} \label{f:Combined_contour2}
\end{figure}

In Figs. \ref{f:Combined_contour2},\ref{f:delta
contour},\ref{f:Newton contour} we show three attractive
three-body potentials that are either pairwise sums of two-body
terms , {\it viz.} Newton's
\begin{equation} \label{Newt}
V_{\rm Newton} = - g \sum_{i<j}^3 \frac{1}{|{\bf x}_{i} - {\bf
x}_{j}|},
\end{equation}
and the $\Delta$-string
\begin{equation}
\label{conf_D} V_{\Delta} = \sigma_{\Delta} \sum_{i<j}^3 |{\bf
x}_{i} - {\bf x}_{j}|,
\end{equation}
or contain such a two-body component in a limited part of the
configuration (shape) space, such as the Y-string
\begin{equation}
\label{conf_Y} V_{Y} = \sigma_{Y} \min_{\bf x}\; \sum_{i=1}^3
|{\bf x_i} - {\bf x}| = \sigma_{Y} \; \sum_{i=1}^3 |{\bf x_i} -
{\bf x_{\rm T}}|,
\end{equation}
where the minimum of the sum occurs at the Torricelli point ${\bf
x} = \bf x_{\rm T}$, see Ref. \cite{taka01}.
\begin{figure}[tbp]
\centerline{\includegraphics[width=2.5in,,keepaspectratio]{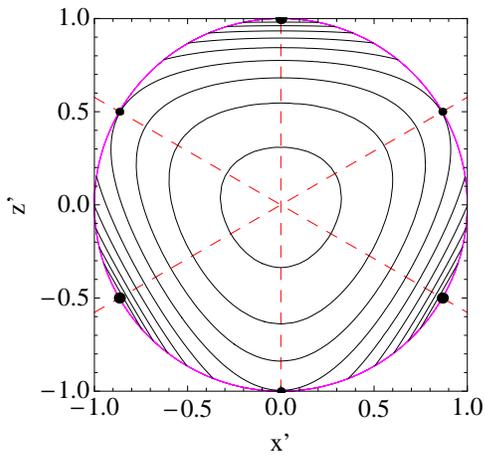}}
\caption{(Color online) Contour plot of the $\Delta$-string
potential as a function of $z^{'}=z=\cos2\chi$ (vertical axis) and
$x^{'}=x \sqrt{1-z^2}=\cos\theta \sin2\chi$ (horizontal axis) for
any fixed value of the hyper-radius $R$. The rest of the legenda is as in Figs. 
\ref{f:Combined_contour} and \ref{f:Combined_contour2}. The center of the circle
is the point with the highest value of the potential.}
\label{f:delta contour}
\end{figure}
\begin{figure}[tbp]
\centerline{\includegraphics[width=2.5in,,keepaspectratio]{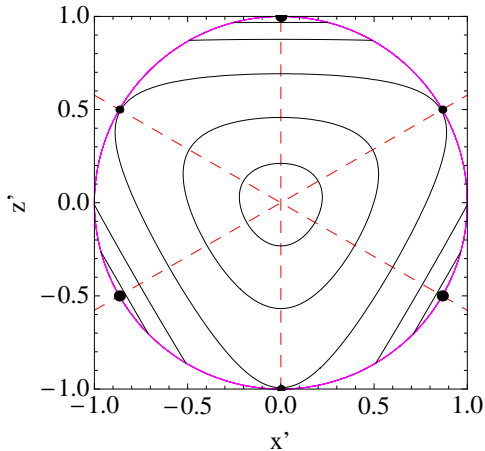}}
\caption{(Color online) Contour plot of the logarithm of the sum
of Newton's two-body potentials as a function of
$z^{'}=z=\cos2\chi$ (vertical axis) and $x^{'} = x \sqrt{1-z^2} =
\cos\theta \sin2\chi$ (horizontal axis) at any fixed value of the
hyper-radius $R$. The rest of the legenda is as in Figs. 
\ref{f:Combined_contour} and \ref{f:Combined_contour2}. As one 
approaches the two-body collision points
(three large solid points on the big circle), the equipotential
contour lines become more and more dense and parallel, finally
reaching infinite density at these points, due to the
singularities/poles present.} \label{f:Newton contour}
\end{figure}
Note that the Y-string potential has perfectly concentric contour
lines within a pear-shaped region of shape space delineated by the
blue dashed line in Fig. \ref{f:Combined_contour2}. As shown in
Ref. \cite{Dmitrasinovic:2009ma}, that ``hyper-rotational" symmetry
leads to a new constant of motion in this part of shape space. A
clear discrimination of the Y-string from the $\Delta$-string
three-quark potentials had been a problem in lattice QCD until
Ref. \cite{Dmitrasinovic:2009ma} showed that the two kinds of
potentials have essentially different hyper-angular dependencies.
The separation of one kind of three-body potential from another is
facilitated by the use of the new variables $r$ and $\phi$. Then
the Y-string component is manifested through the sole dependence
on $r$, whereas the $\Delta$-string is manifested through the
dependence of the potential on the hyper-angle $\phi$, within the
confines of the ``central potential" boundary in terms of ``old"
variables $(\chi,\theta)$) \cite{Dmitrasinovic:2009dy}.

Note that all three potentials in Figs.
\ref{f:Combined_contour2},\ref{f:delta contour},\ref{f:Newton
contour} have essentially (topologically) the same form in the two
hyper-angular (``shape space") variables:
\begin{eqnarray}
V(r,\phi) &=& V(r) + {\delta} V(r) \cos(3\phi) + \ldots
\label{e:V1}. \
\end{eqnarray}
This is a consequence of their permutation symmetry. Any
attractive permutation symmetric potential has its highest value
at the center of the circle ($r=0$)and it decreases monotonically
as one moves radially towards the $r=1$ circle. Moreover, a
permutation symmetric potential is circularly symmetric at the
center (${\delta} V(0) = 0$) and is increasingly broken by a
periodic $\phi$ angular (``two-body") component ${\delta} V(r)
\cos(3\phi)$ as one moves radially towards the $r=1$ circle.

As a consequence of this ``topological equivalence" these
potentials lead to certain kinds of orbits, such as the ``figure
eight" one, that are essentially identical. The hyper-radial part
of the potential does not seem to be very important, so long as it
is attractive, because the stability of the orbit is ensured by
the approximate (dynamical) O(2) symmetry of these potentials. The
details of these potential differ, of course, and therefore lead
to different detailed properties of the amplitude and phase
variations, but not so for the qualitative properties of the
motion. Indeed, if the potential does not contain the periodic
$\phi$-dependent ``two-body" component near the ``outer edge" of
the shape space circle (or near the equator of the shape space
hemisphere), then there is no ``figure-eight" orbit in that
potential.

\subsection{Approximate dynamical O(2) symmetry}
\label{s:dyn symm}

The sum of Newton's or $\Delta$-string two-body potentials is
approximately symmetric under infinitesimal rotations in the shape
space, at least in the central ($r \simeq 0$) part of the ``shape
space", as can be seen in Figs. \ref{f:delta contour},
\ref{f:Newton contour}, whereas the Y-string is exactly symmetric
in the same region, see Fig. \ref{f:Combined_contour2}.

Of course, the two-body potentials are exactly invariant under the
finite (``kinematic") rotations through $\phi = \pm
\frac{2\pi}{3}$, that correspond to cyclic permutations, as well
as under reflections about the three symmetry axes, that
correspond to binary/two-body permutations (``transpositions").

Independence of the potential on the variable $\phi$ is equivalent
to its invariance under (infinitesimal) ``kinematic rotation" O(2)
transformations
\begin{eqnarray}
\delta x^{'} &=& ~~ 2 \varepsilon z^{'} \\
\delta z^{'} &=& - 2 \varepsilon x^{'} , \label{e:x'z' O2 trf} \
\end{eqnarray}
or, in terms of the original Jacobi variables,
\begin{eqnarray}
\delta {\bm \rho} &=& ~~ \varepsilon {\bm \lambda} \\
\delta {\bm \lambda} &=& - \varepsilon {\bm \rho} . \label{e:O2
trf} \
\end{eqnarray}
in the six-dimensional hyper-space. This invariance leads to the
new integral of motion $G_3 = \frac12 \left({\bf p_{\rho}} \cdot
{\bm \lambda} - {\bf p_{\lambda}} \cdot {\bm \rho} \right)$,
associated with the dynamical symmetry (Lie) group $O(2)$ that is
a subgroup of the (full hyper-spherical) $O(6)$ Lie group.

This O(2) symmetry transformation is an infinitesimal version of
the so-called ``kinematic rotations", see Ref.
\cite{Smith:1959zz}, that operate in two, ordinarily different
planes at the same time\footnote{An ordinary space rotation
rotates both position and velocity vectors about the same axis and
through the same angle.}: 1) in the plane of Jacobi vectors ${\bm
\rho}, {\bm \lambda}$, and 2) in the plane of Jacobi momenta ${\bf
p}_{\rho}-{\bf p}_{\lambda}$ (these two planes need not coincide
in general). It is only in the special case of planar motions that
these two planes coincide, and it is only in the (even more
special) case of vanishing (total) angular momentum that the new
constant of motion has presently discernible consequences.

In the case of the sum of two-body potentials, such as the
Newtonian gravity, or the $\Delta$ string potential, this
generalized hyper-angular momentum $G_3$ is not an exact integral
of motion, but an approximate one. The precise consequences of
such an approximate symmetry depend on the initial conditions of
motion, as we shall see below.

\section{Figure-eight motion}
\label{s:figure 8}

\begin{figure}[tbp]
\centerline{\includegraphics[width=3.25in,,keepaspectratio]{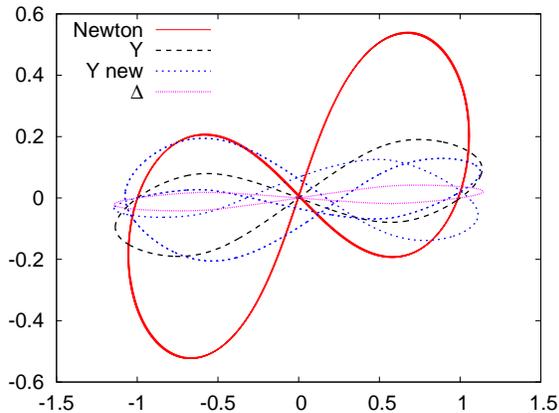}}
\caption{(Color online) Real space trajectories of the
figure-eight and one new solution that passes through the
figure-eight initial configuration for three different potentials.
The legenda are explicitly shown in the upper left-hand corner of
the figure: 1) Newtonian potential figure-eight: red (dark gray
solid) curve; 2) Y-string potential figure-eight: (dark gray
long dashed) curve; 3) Y-string potential new solution: blue
(gray medium-length dashed) curve; 4) $\Delta$-string potential
figure-eight: magenta (light gray short dashed) curve.} \label{f:fig5}
\end{figure}

A periodic ``figure eight" orbit, Fig. \ref{f:fig5}, with
vanishing total angular momentum ($L$=0) has been found by Moore
\cite{Moore1993} in the case of equal masses and gravitational
potential.
It was shown in Ref. \cite{Fujiwara2003b} that the hyper-radius
$R$ is close to being constant along the figure-eight trajectory:
it only makes small-amplitude oscillations in lockstep, i.e. with
the same frequency and locked in phase, with the area of the
triangle and the hyper-angle $\phi$, see below.

\subsection{Time dependence of the hyper-angular motion}
\label{s:hyperangle time dependence}

We take $(r_{\rm init.}, \phi_{\rm init.})=(1, \frac13 \pi)$ as
the initial condition, which is one of three identical
configurations, up to permutations. This is a collinear
configuration with one particle in the middle of the other two.
The initial velocities are such that the total angular momentum
vanishes.

\subsubsection{Newtonian gravity}
\label{s:Newton2}

\begin{figure}[tbp]
\centerline{\includegraphics[width=3.25in,,keepaspectratio]{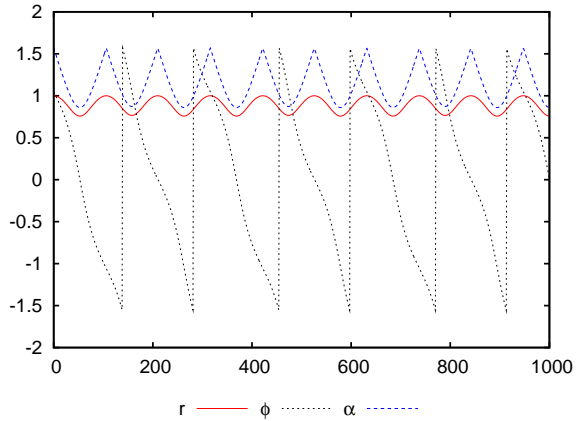}}
\caption{(Color online) The time dependence of the hyper-angular
radius $r$ red (solid), and the hyper-angles $\alpha = \sin^{-1}r$
blue (long dashed) and $\phi$ gray (short dashed) of the
figure-eight solution in Newton's potential. The legenda are
explicitly shown below the figure.} \label{f:Newton fig6}
\end{figure}

In Fig. \ref{f:Newton fig6}
we see that both hyper-angular variables $(r,\phi)$ oscillate with
the same frequency and locked in phase along the figure-eight
trajectory.

\subsubsection{The Y- and $\Delta$ string potentials}
\label{s:strings2}

\begin{figure}[tbp]
\centerline{\includegraphics[width=3.25in,,keepaspectratio]{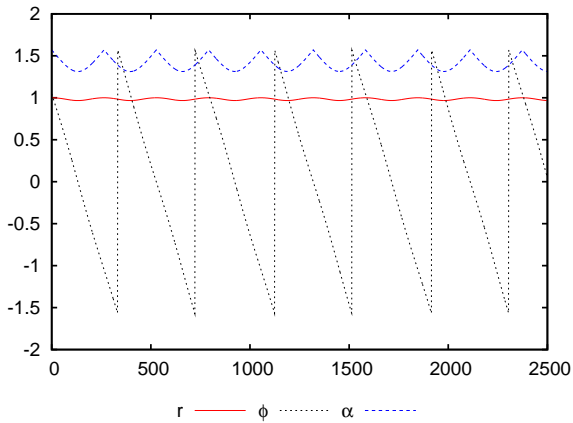}}
\caption{(Color online) The time dependence of the hyper-angular radius $r$ red
(solid), and the hyper-angles $\alpha = \sin^{-1}r$ blue (long
dashed) and $\phi$ gray (short dashed) of the figure-eight
solution in the Y-string potential. The legenda are
explicitly shown below the figure.} \label{f:Y fig7}
\end{figure}

\begin{figure}[tbp]
\centerline{\includegraphics[width=3.25in,,keepaspectratio]{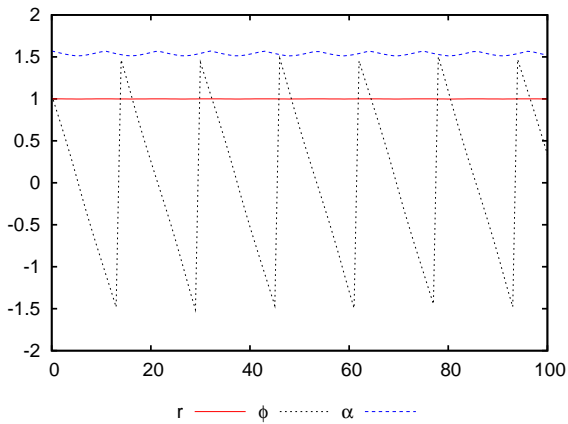}}
\caption{(Color online) The time dependence of the hyper-angular
radius $r$ red (solid), and the hyper-angles $\alpha = \sin^{-1}r$
blue (long dashed) and $\phi$ gray (short dashed) of the
figure-eight solution in the $\Delta$-string potential. The legenda are
explicitly shown below the figure.}
\label{f:Delta fig8a}
\end{figure}

A similar situation is present in the other two potentials: the
hyper-radius $R(t)$ is almost constant along this trajectory: it
makes small-amplitude oscillations in phase with the area of the
triangle, Fig. \ref{f:Y fig7}.
Similarly for the $\Delta$ string, see Fig. \ref{f:Delta fig8a}.

\subsection{Hyper-angular $\phi$ dependence}
\label{s:hyperangular dependence}

\begin{figure}[tbp]
\centerline{\includegraphics[width=3.25in,,keepaspectratio]{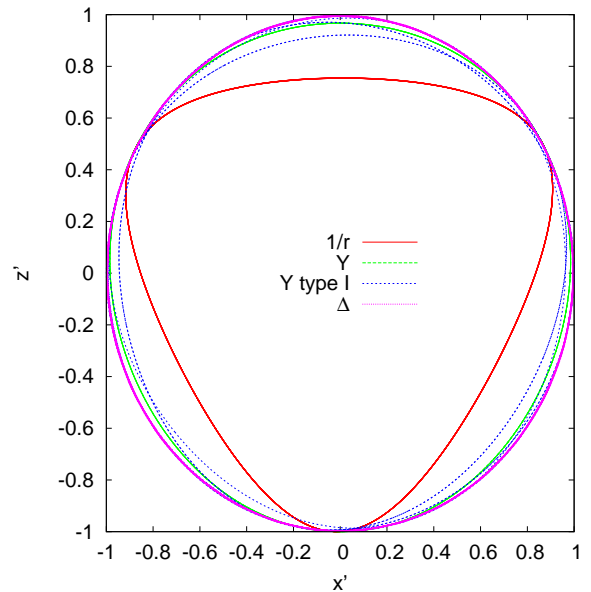}}
\caption{(Color online) Trajectories of the figure-eight and one
new solution in three different potentials in terms of $z^{'}=z=\cos2\chi$ 
(vertical axis) and $x^{'} = x \sqrt{1-z^2} =
\cos\theta \sin2\chi$ (horizontal axis). The legenda are explicitly shown in the middle of
the figure: 1) Newtonian potential figure-eight: red (dark gray
solid) curve; 2) Y-string potential figure-eight: (dark gray
long dashed) curve; 3) Y-string potential new solution: blue
(gray medium-length dashed) curve; 4) $\Delta$-string potential
figure-eight: magenta (light gray short dashed) curve; 5) unit circle: magenta 
(dark gray solid).} \label{f:alpha phi}
\end{figure}

One can see in Fig. \ref{f:alpha phi} that the periodicity of the
figure-eight motion is determined by the braiding angle $\phi$.
Here one can also see that the actual path in the shape space,
Fig. \ref{f:alpha phi}, taken by the Newtonian three-body system
is remarkably close to the Newtonian iso-potential lines in Fig.
\ref{f:Newton contour}. If this were true, then the hyper-radius
would be constant along the figure-eight orbit, but it is not:
$R(t)$ and $r(t)$ are periodic functions, with the same basic
frequency of $3 \phi(t)$ and locked in phase, i.e. they oscillate
about their average values as follows
\begin{eqnarray}
\phi &=& \langle \dot{\phi} \rangle t + {\delta \phi}\sin(3\phi) +
\ldots
\nonumber \\
r(\phi) &=& \overline{r} + {\delta r}\sin(3\phi) + \ldots
\nonumber \\
R(\phi) &=& \overline{R} + {\delta R}\sin(3\phi) + \ldots
\label{e:G1}. \
\end{eqnarray}
This phase- and frequency locking provide an important constraint
that effectively reduces the number of independent degrees of
freedom to two. In other words $\phi$ is the (approximate) cyclic,
or ``ignorable" variable of the figure-eight periodic motion that
may be integrated out/ignored/. The conjugate action variable
$G_3$ is the associated (approximate) integral of motion.

\subsection{The action variable conjugate to $\phi$}
\label{s:action}

In order to find the appropriate action variable we look at the
(``kinematic") hyper-angular momentum $G_3$ that reads
\begin{eqnarray}
G_3 &=& \frac{m}{4} \left(R r \right)^2 {\dot \phi} = \frac{m}{4}
\left(R \sin\alpha \right)^2 {\dot \phi} \label{e:G2} \
\end{eqnarray}
(with vanishing angular momentum $L=0$) as a function of
permutation symmetric variables $R, r=\sin\alpha$ and $\phi$, and
oscillates as a periodic function of the hyper-angle $3\phi$.
Hence it follows that $G_3$ and ${\dot \phi}$ must be (almost)
constant in orbits with $L=0$, $R\simeq$const and $r \simeq 1$. In
other words, the angle $\phi$ grows (or decreases, depending on
the orientation of the motion) almost linearly in time, which is
confirmed by our numerical results.

The time/hyper-angle average $\overline{G}_3$
\begin{eqnarray}
\overline{G}_3 &=& \frac{1}{T}\int_{0}^{T}{G}_3~ dt  =
\frac{1}{T}\int_{0}^{T} \frac{m}{4} \left(R r \right)^2
{\dot \phi}~ dt \nonumber \\
&=& \frac{1}{2 \pi}\int_{0}^{2\pi} \frac{m}{4} \left(R r \right)^2
~d\phi \label{e:G3} \
\end{eqnarray}
is a non-vanishing constant on the figure-eight orbit, furnishing
the (approximate) action variable that goes together with the
(linearized) hyper-angle $\phi$ for this periodic motion. The
approximate constancy of ${G}_3 \simeq \overline{G}_3 \neq 0$ is
the cause of dynamical stability of the figure-eight orbit: The
vanishing angular momentum ($L=0$) three-body relative motion
kinetic energy
\begin{eqnarray}
T_{\rm kin} &=& \frac{m}{2} \left(\dot{R}^2 +
\left(\frac{R}{2}\right)^2 \left(\frac{\dot{r}^2}{1 - r^2} +
\left(r\,\dot{\phi}\right)^2
\right) \right]  \nonumber \\
&=& \frac{m}{2} \left[\dot{R}^2 + \left(\frac{R}{2}\right)^2
\left({\dot{\alpha}^2} + \left(\dot{\phi}\, \sin\alpha\right)^2
\right) \right] \label{e:kin} \
\end{eqnarray}
has the form of the single-particle kinetic energy in polar
coordinates, albeit with polar radius $\frac{R}{2}$ reduced by
half in the ``hyper-angular kinetic energy" $\frac{m}{2}
\left(\frac{R}{2}\right)^2 \left[{\dot{\alpha}^2} +
\left(\dot{\phi}\, \sin\alpha\right)^2 \right]$. This means that
another (hyper-) angular momentum-like three vector ${\bf G}$ is
conserved when the potential does not depend on the two angles
$(\alpha ,\,{\phi})$.

In the case when the potential depends on $\alpha$, but does not
depend on ${\phi}$, or has only small variations with ${\phi}$,
then the ``azimuthal hyper-angular momentum" $G_3 = \frac{\partial
T_{\rm kin}}{\partial {\dot{\phi}}}$ is approximately constant:
$$\dot{G}_3 = - \frac{\partial V(r, \phi)}{\partial \phi} = 3
{\delta} V(r) \sin(3\phi) +\ldots ~.$$ Then $G_3 \simeq
\overline{G}_3 \neq 0$ provides a repulsive term $\frac{2 G_3^2}{m
R^2}$ in the effective hyper-radial potential $V_{\rm eff}(R) =
\frac{2 G_3^2}{m R^2} + V_{\rm 3-body}(R)$ that prevents the
system from collapsing to a point, just as the (ordinary) angular
momentum $L \neq 0$ does in the two-body problem.

\subsubsection{Newtonian gravity}
\label{s:Newton1}

The temporal variation of the hyper-radius $R(t)$ and the
hyper-angular momentum $G(t)$ in the Newtonian gravitational
potential are shown in Fig. \ref{f:Newton fig10}.
\begin{figure}[tbp]
\centerline{\includegraphics[width=3.25in,,keepaspectratio]{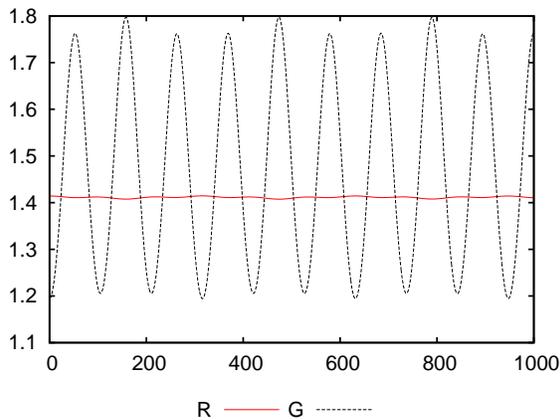}}
\caption{(Color online) The time dependence of the hyper-radius $R$ red (solid), 
and the hyper-angular momentum $G$ gray (dashed) of the
figure-eight solution in Newton's potential. The legenda are
explicitly shown below the figure.} \label{f:Newton
fig10}
\end{figure}

\subsubsection{The Y- and $\Delta$ string potentials}
\label{s:strings1}

\begin{figure}[tbp]
\centerline{\includegraphics[width=3.25in,,keepaspectratio]{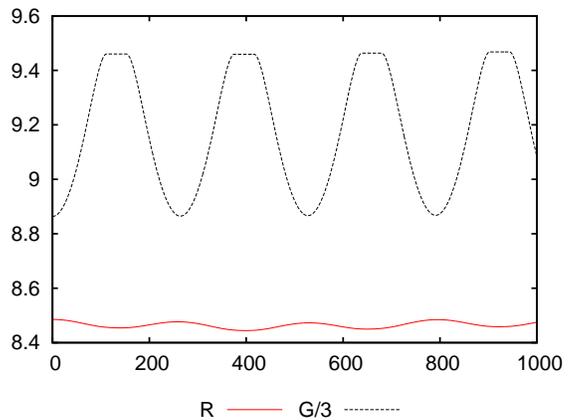}}
\caption{(Color online) The time dependence of the hyper-radius
$R$ red (solid), and one third of the hyper-angular momentum
$G/3$ gray (dashed) of the figure-eight solution in the Y-string
potential. The legenda are explicitly shown below the figure.} \label{f:Y fig11}
\end{figure}

The temporal variation of the hyper-angular momentum $G(t)$ and
$R(t)$, the former reduced by factor three, so as to emphasize the
small variation of $R$ in the Y-string potential are shown in Fig. \ref{f:Y fig11}.
Note the cut off peaks of the sines (``flat tops") of the
hyper-angular momentum $G/3$ due to the exact dynamical O(2)
symmetry in that part of the configuration space.

\section{New Y-string periodic orbits}
\label{s:new sol}

Note, moreover, that the permutation symmetric three-body
potential $V_{\rm 3-body}(\sin \alpha,\phi)$ in the region of the
figure-eight orbit (i.e. on the outer fringes of the $(r = \sin
\alpha,\phi)$ circle) is attractive as a function of $r = \sin
\alpha$, with a minimum at the unit circle ($r=1$), and a strictly
periodic function of the (triple) hyper-angle $3 \phi$.

Then it should be no surprise that the figure-eight motion of the
three-body system, with its almost constant hyper-radius $R$, has
many similarities with that of the spherical pendulum in an
inhomogeneous (azimuthally periodic) gravitational field:
figure-eight orbit corresponds to rotations, but there are other
qualitatively different kinds of motions that we shall display and
briefly discuss in this section.

There is a small, yet pronounced non-linearity in the figure-eight
motion's $\phi$'s temporal dependence, particularly near the $\phi
= 0, \pm \frac{2\pi}{3}$ points. These three points/lines in the
$(r,\phi)$ circle, see Fig. \ref{f:alpha phi}, correspond to the
configurations of closest two-body approach in real space. Of
course, the figure-eight orbit does not touch the ``unit circle"
at these three values of $\phi$, so there are no two-body
collisions in this type of orbit.

Yet, this suggests that there might be other, perhaps multiply
periodic solutions with ``trajectories" in the $(r,\phi)$ plane
that touch the $r$=1 circle at the hyper-angle values other than
$\phi = \pm \frac{\pi}{3}, {\pi}$ and/or approach the unit circle
(the equator of the shape hemisphere) even closer to the two-body
collision points. The latter fact means that the corresponding
trajectories in real space are ``narrower" than the figure-eight
one.

We have studied this region more closely and found several new
periodic solutions with lesser symmetry than the figure-eight one
that pass through the ``infinitesimal" neighborhood of the initial
state, but only in the Y-string and the $\Delta$-string potentials
(i.e. not in Newton's gravity, as yet). We display two interesting
new orbits below. The initial conditions are given in Table
\ref{tab:init cond}.

\begin{table}[tbh]
\begin{center}
\caption{(Color online) The initial conditions for the solutions
shown in this paper: $d$ is the value of the initial distance
between the outer left, or right particle and the middle one, and
the velocities (both are in dimensionless units where the masses
and the coupling constants have been set equal to unity); the
angle $\theta$ is in radians.}
\begin{tabular}{ccccc}
\hline \hline name & $d$ & $v$ & $\theta$(rad) & ${\rm potential}$\\
\hline
fig.8 & 6 & 1.37 & 1.205 & Y-string \\
type I & 6 & 1.32 & 1.437 & Y-string \\
type II & 6 & 4.53 & 1.40 & Y-string \\
fig.8 & 1 & 0.6355 & 0.5736 & Newton \\
fig.8 & 1 & 0.536 & 1.49287 & $\Delta$-string \\
\hline \hline
\end{tabular}
\label{tab:init cond}
\end{center}
\end{table}

\subsection{Type I (``linear-in-$\phi$") reduced symmetry solution}
\label{s:typeI}

\begin{figure}[tbp]
\centerline{\includegraphics[width=3.25in,,keepaspectratio]{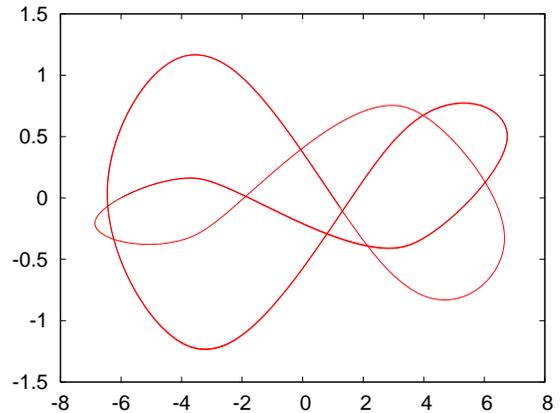}}
\caption{(Color online) Real space trajectory of particle number 1 of the type-I new
solution that passes through the figure-eight initial
configuration. The trajectory of particle number 3 is a reflection
about the line dividing this trajectory vertically.}
\label{f:double12}
\end{figure}

\begin{figure}[tbp]
\centerline{\includegraphics[width=3.25in,,keepaspectratio]{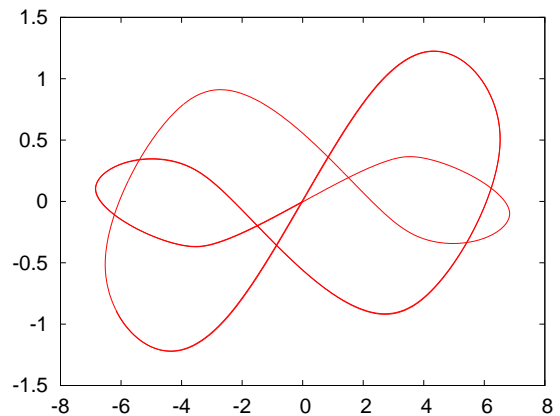}}
\caption{(Color online) Real space trajectory of particle number 2 of the
type-I new solution that passes through the figure-eight initial
configuration.} \label{f:double13}
\end{figure}

First note that the real-space trajectory of (``right-hand-side")
particle number 2 in Fig. \ref{f:double13} is different from the
one of the (``central") particle number 1 in Fig. \ref{f:double12}
thus making it clear that this is a periodic, but not a
choreographic motion.
In other words, this solution is symmetric ``merely" under the
two-body permutation group $s_2$, rather than under the three-body
permutation group $s_3$. Due to the reduced symmetry, one particle
executes an ``independent" motion, whereas the other two move on
orbits that are mirror images of each other, very much like those
in Ref. \cite{Broucke2006}. This means that this new solution is
probably a bifurcation of the figure-eight orbit as a function of
particle masses related to those found in the Ref.
\cite{Galán2002}, i.e. as a function of explicit $s_3$ permutation
symmetry breaking.

The figure-eight orbit touches the unit circle at three points of
the equilateral triangle defined by $\phi = \pm \frac13 \pi, \pi$,
see Fig. \ref{f:alpha phi}, whereas this new solution touches it
at only one vertex of this equilateral triangle {\it viz.} $\phi =
\frac13 \pi$, and ``cuts corners" at the other two, only to touch
the unit circle at four other values of $\phi$ that are different
from the two-body collisions points $0, \pm \frac23 \pi$. This is
still a periodic solution with a period of $8 \pi$, i.e. it takes
four cycles of the hyper-angle $\phi$ to complete one period, but
with several different hyper-angular frequencies, instead of the
single basic frequency $3 \phi$. This fact may not be immediately
visible to the naked eye, as these frequencies are close to $3
\phi$, but shows up as ``beats" in the time dependence of the
amplitudes.

\begin{figure}[tbp]
\centerline{\includegraphics[width=3.25in,,keepaspectratio]{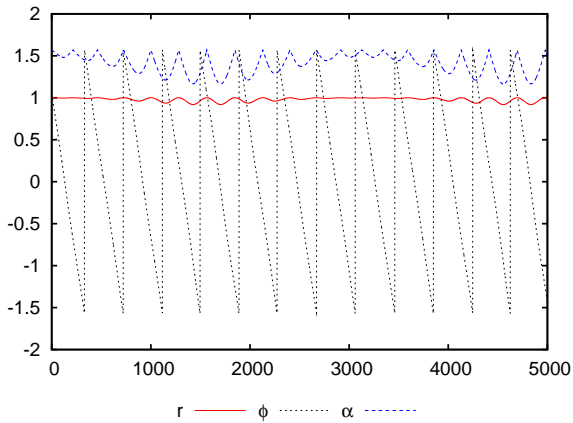}}
\caption{(Color online) The time dependence of the hyper-angular
radius $r$ red (solid), and the hyper-angles $\alpha = \sin^{-1}r$
blue (long dashed) and $\phi$ gray (short dashed) of the type-I
new solution in the Y-string potential that passes through the
figure-eight initial configuration. The legenda are
explicitly shown below the figure. Note that $\phi$ moving from 0
to $2 \pi$ corresponds to two segments between vertical
lines/discontinuities due to the numerical evaluation of inverse
trigonometric functions. Note that one complete period of the
motion corresponds to eight such segments, i.e. to $\phi$ changing
from 0 to $8 \pi$, or to four complete revolutions around the
$(r,\phi)$ circle.} \label{f:Y new 14}
\end{figure}

\begin{figure}[tbp]
\centerline{\includegraphics[width=3.25in,,keepaspectratio]{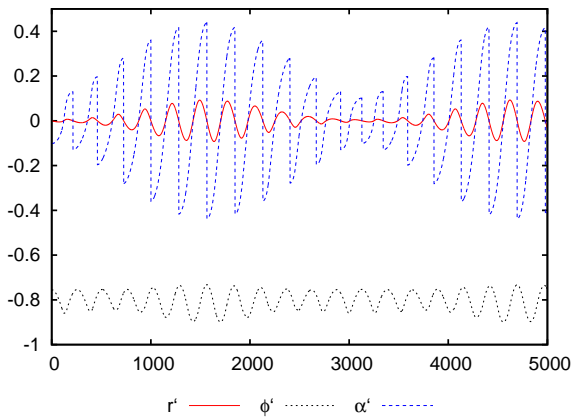}}
\caption{(Color online) The first derivatives of the time
dependence of the hyper-angular radius $\dot{r}$ red (solid), and
the hyper-angles $\dot{\alpha} = \frac{d}{dt} \sin^{-1}r$ blue
(long dashed) and $\dot{\phi}$ gray (short dashed) of the type-I
new solution in the Y-string potential that passes through the
figure-eight initial configuration.} \label{f:Y new 15}
\end{figure}

The hyper-angle $\phi$ in this solution (still) grows (or
descends) indefinitely, so this solution also corresponds to a
kind of rotation of the pendulum, but with a changing angular
velocity, see Fig. \ref{f:Y new 14}.
The time derivatives show the beats more clearly, see Fig.
\ref{f:Y new 15}.
The temporal variation of the hyper-radius $R(t)$ is shown in Fig.
\ref{f:Y new 16}, and that
\begin{figure}[tbp]
\centerline{\includegraphics[width=3.25in,,keepaspectratio]{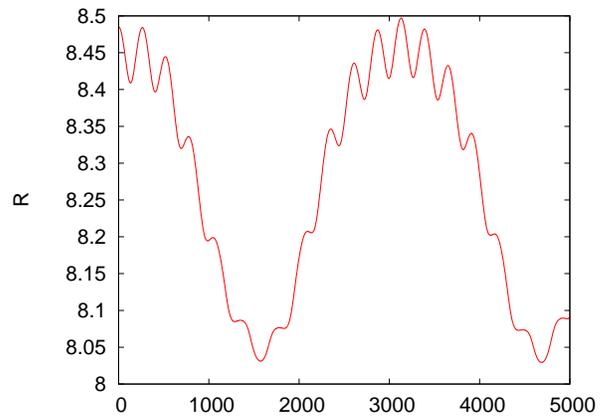}}
\caption{(Color online) The time dependence of the hyper-radius
$R$ of the type-I new solution in the Y-string potential that
passes through the figure-eight initial configuration.} \label{f:Y
new 16}
\end{figure}
of the hyper-angular momentum $G(t)$ is shown in Fig. \ref{f:Y new
17}.
\begin{figure}[tbp]
\centerline{\includegraphics[width=3.25in,,keepaspectratio]{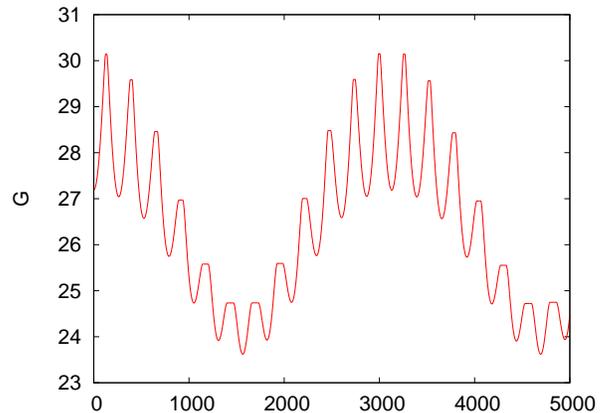}}
\caption{(Color online) The time dependence of the hyper-angular
momentum $G$ of the type-I new solution in the Y-string potential
that passes through the figure-eight initial configuration. Note
the cut off peaks of the sines (``flat tops").} \label{f:Y new 17}
\end{figure}
Note the beats in the time evolution of $R(t)$ and $G(t)$, as
advertised.

\subsection{Type II (``oscillating $\phi$") reduced symmetry solution}
\label{s:typeII}

\begin{figure}[tbp]
\centerline{\includegraphics[width=3.25in,,keepaspectratio]{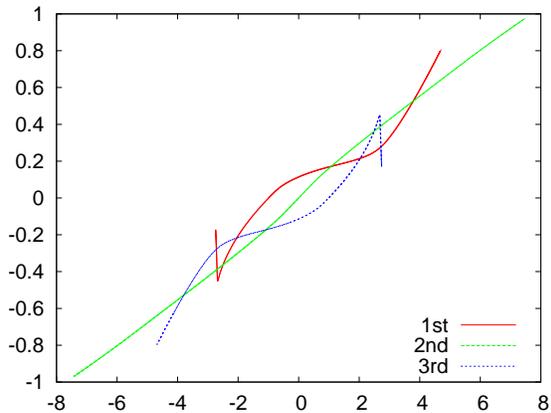}}
\caption{(Color online) Real space trajectories of particle number 1 red
(solid), particle number 2 green (light gray dashed) and particle number 
3 blue (dark gray dashed) line in the type-II new solution that 
passes through the figure-eight initial configuration. The legenda 
are explicitly shown in the lower right-hand side corner of
the figure.} \label{f:Y new 2a1}
\end{figure}

\begin{figure}[tbp]
\centerline{\includegraphics[width=3.25in,,keepaspectratio]{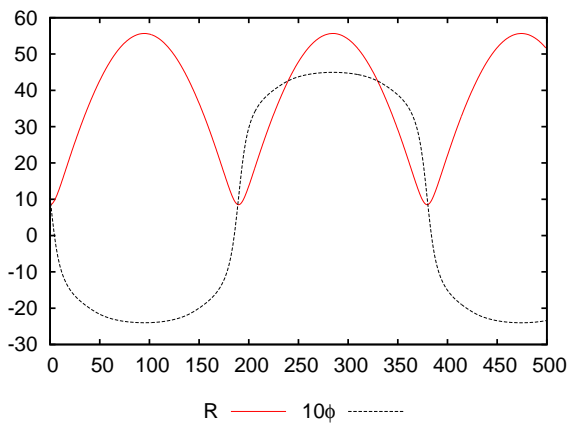}}
\caption{(Color online) The time dependence of the hyper-radius
$R$ red (solid), and 10$\times \phi$ gray (short dashed) in the
type-II ``oscillating" solution in the Y-string potential. The 
legenda are explicitly shown below the figure.}
\label{f:Y new 2c}
\end{figure}

First note that the real-space trajectory of particle number 2 in
this solution, Fig. \ref{f:Y new 2a1}, is different from the one
of particle number 1, thus making it clear that this is also a
periodic, but not a choreographic motion. The trajectory of
particle number 3 (blue on-line) is a reflection of trajectory of
particle number 1 about the origin.
At first, the aforementioned action-angle variables do not seem
appropriate for this new periodic orbit, indeed the (formerly)
linear increase/decrease of the hyper-angle $\phi$ is now subject
to substantial modifications: after initial rapid (hyper-)
rotation in the clock-wise direction starting from $\phi = 0$, it
slows down and stops around $\phi_{\rm min} \simeq - 0.764 \pi$,
Fig. \ref{f:Y new 2c}, then changes the direction of motion and
swings back yet again only to stop, this time around $\phi_{\rm
max} \simeq 1.431\pi$, then repeating this cycle ad infinitum.
Note that the maximal difference (twice the amplitude) of $\phi$
is numerically close to being a simple fraction of $\pi$, i.e.
$\Delta \phi = \phi_{\rm max} - \phi_{\rm min} = 13.0001
\frac{1}{6}\pi$, whereas the average value ${\overline \phi} =
\frac12 (\phi_{\rm max} + \phi_{\rm min})$ is numerically close to
$\frac{1}{3}\pi$. We suspect that $\frac{13}{6}\pi$ and
$\frac{1}{3}\pi$ are the exact values and that the deviations from
our numerical values are due to rounding-off errors.
This resembles the oscillations of a variable-length pendulum.
Indeed, Fig.\ref{f:Y new 2c} shows that the hyper-radius $R(t)$ is
oscillating with the same frequency and phase as the hyper-angle
$\phi(t)$, thus extending the analogy with the variable-length
pendulum model.

\section{Conclusions}
\label{s:concl}

We have studied the figure-eight motion in three different
three-body potentials in terms of permutation symmetric variables
$(R, r=\sin\alpha)$, and the braiding hyper-angle $\phi$. The
existence of this orbit depends on the periodic dependence of the
potential on the braiding hyper-angle $\phi$ that ``guides" the
figure-eight orbit(s) around the two-body collision points.

The figure-eight orbits in the triangle shape space are generally
close to their iso-potential lines, though formal arguments show
that they cannot be exactly identical \cite{Chen2001}. Thus the
exact analytic solutions ought to be sought among (small)
oscillations about the iso-potential lines, with basic frequency
$3 \phi$.

The Hamiltonian of three identical particles in a permutation
symmetric potential with vanishing total angular momentum has
certain similarities with that of a spherical pendulum in
inhomogeneous azimuthally-periodic potentials, which, in turn,
suggests existence of other types of solutions.

We have found two new periodic solutions in the Y-string potential
that pass through the figure-eight initial state, but do not share
its symmetry. One of these solutions (type I) has a monotonically
rising/descending hyper-angle $\phi$, just like the figure-eight
orbit, but a different pattern of syzygies, whereas the second
(type II) new solution's $\phi$ is oscillating about its average
value of $\frac{\pi}{3}$, with the hyper-radius $R$ following
suit.

All of these orbits are clearly characterized by their $(R, r,
\phi)$ behaviors that display certain similarities, despite their
independent and seemingly random form of trajectories in the
configuration space. Thus we believe this to be a good set of
variables to mathematically simplify and describe all periodic
orbits of three identical bodies.

A few words about the history of this subject and our approach to
it might be in order now. In our study Ref.
\cite{Dmitrasinovic:2009ma} of the so-called Y-junction and the
$\Delta$-string potentials we found an integral of three-body
motion, when the three-body potential depends on only two, rather
than three, independent three-body variables, {\it viz.} the
hyper-radius (or the moment of inertia divided by the
quark/particle mass) and the area of the triangle defined by the
three bodies \footnote{The exact Y-string potential does not
always conserve this new integral of motion, due to its
angle-dependent ``two-body" parts, but is valid in the major part
of the ``triangle shape space" and will be shown to play a role in
the existence of a new, figure-eight-shaped closed trajectory.}.
As the static three-body potential may depend on (at most) three
independent scalar variables, our observation naturally begged the
question: what is the third independent three-body variable in
this set? \footnote{As the first two variables (the hyper-radius
and the area of the triangle) are manifestly invariant under
permutations of the three particles, we call this set
``permutation symmetric".} It was an attempt to answer this
question that brought us to the present permutation symmetric
variables. We are not the first ones to use them, however:
Chenciner and Montgomery have used these variables (these authors
call $\theta$ ``our" variable $\phi$) to parametrize the triangle
shape space in Ref. \cite{Chenciner2000}. According to Ref.
\cite{Chenciner2000}, H. Hopf was the first one to introduce these
variables, Ref. \cite{Hopf 1931}.

\noindent{\bf Acknowledgments}

This work was financed by the Serbian Ministry of Science and
Technological Development under grant number 141025. One of us
(V.D.) wishes to thank A. Ohnishi for hospitality at the Yukawa
Institute for Theoretical Physics, Kyoto, where some of this work
was carried out. The same author thanks T. Sato and H. Suganuma
for valuable conversations.

\end{document}